\documentclass[twocolumn,showpacs,preprintnumbers,amsmath,amssymb]{revtex4}

\usepackage{psfig}
\usepackage{graphicx}
\usepackage{dcolumn}
\usepackage{bm}

\begin{document}


\title{Systematic study of carrier correlations in the electron-hole recombination dynamics of quantum dots}

\author{T. Berstermann}
\author{T. Auer}
\author{H. Kurtze}
\author{M. Schwab}
\author{D.R. Yakovlev}
\author{M. Bayer}
\affiliation{Experimentelle Physik II,
             Universit\"at Dortmund,
             44221 Dortmund, Germany}

\author{J. Wiersig}
\author{C. Gies}
\author{F. Jahnke}
\affiliation{Institute for Theoretical Physics,
           University of Bremen,
           28334 Bremen, Germany}

\author{D. Reuter}
\author{A.D. Wieck}
\affiliation{Angewandte Festk\"orperphysik,
             Ruhr-Universit\"at Bochum,
             D-44780 Bochum,
             Germany}

\date{\today}

\begin{abstract}
The ground state carrier dynamics in self-assembled (In,Ga)As/GaAs
quantum dots has been studied using time-resolved
photoluminescence {\sl and} transmission. By varying the dot
design with respect to confinement and doping, the dynamics is
shown to follow in general a non-exponential decay. Only for
specific conditions in regard to optical excitation and carrier
population, for example, the decay can be well described by a
mono-exponential form. For resonant excitation of the ground state
transition a strong shortening of the luminescence decay time is
observed as compared to the non-resonant case. The results are
consistent with a microscopic theory that accounts for deviations
from a simple two-level picture.
\end{abstract}

\maketitle

\section{Introduction}

The carrier recombination dynamics in semiconductor quantum dots
(QDs) is typically analyzed in terms of exponential decays with a
characteristic time constant $\tau$ in which all possible decay
channels are comprised by adding the corresponding decay rates to
give the total decay rate $1 / \tau$. This kind of decay has been
adopted from experiments in atomic physics which are
discussed using two-level schemes corresponding to the following
scenario: An electron has been excited to a higher lying atom
shell, from which it relaxes to a vacancy of a lower lying shell.
In many cases the relaxation is dominated by radiative
recombination, for which mono-exponential decays give an
appropriate description of the relaxation process.

Due to the three-dimensional confinement of carriers,
semiconductor QDs resemble the solid state analogue of atoms. This
has been underlined by the demonstration of effects observed
before in atom optics such as a radiatively limited spectral line
width \cite{Borri2004}, antibunching in the single photon emission
statistics \cite{Michler2000}, a square-root power broadening for
resonant excitation \cite{Stufler2005} etc. Most of these results
were obtained at cryogenic temperatures. At elevated temperatures
the scattering of confined carriers with lattice phonons, for
example, becomes strong, as manifested by a strong broadening of
the optical transitions \cite{T-dependence}. To some extent, this
broadening resembles the collision induced broadening of optical
transitions in high pressure atom gases.

Furthermore, experiments addressing electron-hole recombination in
semiconductors are often performed in a way that not only two
electronic levels are involved. Instead, a pulsed laser excites
carriers non-resonantly above the barrier, from where they are
captured by the confinement potential and relax towards the QD
ground state. This situation can be thought to be analogous to a
situation in which the atoms have been ionized to a plasma of
electrons and ions. During plasma cooling, the electrons are
trapped by the ions and relax by photon emission.

Under such conditions the carrier dynamics can in general not be
described by a mono-exponential decay, in agreement with many
observations reported in literature for QD ensembles. On the other
hand, there have been also reports about exponential decays in
studies of such ensembles \cite{exponential}. Also for single QD
experiments indications for a non-exponential dynamics have been
found \cite{Hours2005}. The observed non-exponentiality has been
ascribed to various origins such as carrier diffusion to the
quantum dots \cite{Schweizer1998,Sermage2003}, state filling
effects due to Pauli blocking
\cite{Samuelson1999,Forchel1996,Merz1996}, inhomogeneities
concerning the electron-hole overlap
\cite{Taylor2003,Lounis2003,Kamenev2005,Bimberg2002,Jiang1995,Yuang1994},
QD potential fluctuations from the quantum confined Stark effect
due to charged defects in the QD vicinity \cite{Bimberg2000} as
well as formation of optically inactive excitons with parallel
electron and hole spins \cite{Lounis2003,Langbein2003}. All these
factors may be of relevance for particular experimental
situations.

However, many studies have been done for specific situations
regarding the QD properties, from which it is hard to develop a
systematic picture. Here we have performed time-resolved studies
of the carrier dynamics covering a wide range of parameters with
respect to these properties such as confinement potential height
and residual carrier population. In addition, the optical
excitation conditions have been chosen such that many of the
factors mentioned above can be ruled out, as described in detail
below. For example, the excitation power was chosen so low, that
multiexciton effects leading to state filling cannot occur. The
influence of carrier diffusion has been ruled out by comparing
excitation above the barrier to excitation below the barrier. By
doing so, also the influence of the environment on the confinement
potential shape has been under control.

In the following we present a detailed study of the dynamics of carriers in
the QD
exciton ground state. We show that decays which are to a good
approximation mono-exponential can occur, but only under very
specific conditions such as fully resonant excitation or very
strong QD confinement. Under other circumstances non-exponential
decays are found. Interestingly, strictly resonant excitation
leads also to a pronounced enhancement of the carrier
recombination rate.

The paper is organized as follows. In the next section we briefly
discuss the theory of QD photoluminescence~\cite{Jahnke2006}, which
is used to analyze the subsequent experimental studies. In
Section~\ref{sec:samples} details of the structures under study
are given together with a description of the experimental
techniques. The experimental data are presented and discussed in
Section~\ref{sec:discussion} and the comparison with the numerical
results is provided in Section~\ref{sec:numerics}.

\section{Theory}
\label{sec:theory}
In our case the dynamics of electrons and holes in QDs was studied
by two different spectroscopic techniques: time-resolved
photoluminescence and time-resolved transmission.
We assume that the carriers quickly loose coherence after their
generation by pulsed laser excitation, e.g. by relaxation, so that
we address only incoherent electron and hole populations.

(i) The intensity $I \left( \omega \right)$ in time-resolved
photoluminescence (TRPL) experiments is given by the temporal
evolution of the number of photons from electron-hole
recombination at the detection frequency $\omega$,
\begin{equation}\label{intensity}
I \left( \omega \right) = \frac{d}{dt} \sum_{\xi} \langle
b_{\xi}^{\dagger} b_{\xi} \rangle\Big|_{|{\bf{q}}| = \omega/c} \ ,
\end{equation}
where $b_{\xi}^{\dagger}$ and $b_{\xi}$ are the creation and
annihilation operators of a photon in state $\xi$, which is given
by the wave vector $\bf{q}$ and the polarization vector. The
brackets $\langle \ldots \rangle$ symbolize the quantum mechanical operator
averages.

(ii) A second, independent method, which allows to draw
conclusions about the dynamics of the electron and hole
populations, is time-resolved differential transmission (TRDT).
The electron and hole populations are described by the expectation
values $f_\nu^e = \langle e_\nu^{\dagger} e_\nu \rangle$ and
$f_\nu^h = \langle h_\nu^{\dagger} h_\nu \rangle$, respectively.
Here, $e_\nu^\dagger$ and $e_\nu$ ($h_\nu^\dagger$ and $h_\nu$)
are the creation and annihilation operators of an electron (hole)
in a state $\nu$, including the QD shell index and the spin
orientation.

In the following, we are interested in the interplay of photon and
population dynamics due to spontaneous recombination,
\begin{eqnarray}\label{photonpopulation}
\frac{d}{dt} \langle b_{\xi}^{\dagger} b_{\xi} \rangle & = &
\hspace{0.25cm}\frac{2}{\hbar} \mbox{Re} \sum_\nu g_{\xi\nu}^* \langle b_{\xi}^{\dagger}
h_\nu e_\nu \rangle \ , \\
\label{carrierpopulation}
\frac{d}{dt} f_\nu^{(e,h)}\Big|_{\mbox{\footnotesize opt}} & = & -\frac{2}{\hbar} \mbox{Re} \sum_\xi g_{\xi\nu}^*\langle b_{\xi}^{\dagger} h_\nu e_\nu
\rangle \ .
\end{eqnarray}
The carrier populations are also subject to carrier-carrier
Coulomb interaction~\cite{Nielsen04} and to carrier-phonon
interaction~\cite{Seebeck05}. The dynamics of both photon and
carrier population are determined by the interband photon-assisted
polarization $\langle b_{\xi}^{\dagger} h_\nu e_\nu \rangle$ and
its  complex conjugate $\langle b_{\xi} e_\nu^{\dagger}
h_\nu^{\dagger} \rangle$. The former describes the emission of a
photon due to the recombination of an electron-hole pair, while
the latter describes the inverse process, the creation of an
electron-hole pair via photon absorption. The strength of the
interband polarization is determined by the coupling matrix
element of the electron-hole transition to the electromagnetic
field, $g_{\xi\nu}$.

For solving Eqs.~(\ref{photonpopulation})
and~(\ref{carrierpopulation}) the interband polarization needs to
be known, which is given by its free evolution, by dephasing, by
excitonic contributions, by stimulated emission (in the case of
QDs embedded into a microcavity~\cite{GWKJ06,Ulrich06}), and by
spontaneous emission, for which
  the source term is
\begin{eqnarray}
i \sum_\alpha g_{\xi\alpha} \langle e_\alpha^{\dagger} e_\nu h_\alpha^{\dagger} h_\nu
\rangle.
\end{eqnarray}
The corresponding equation of motion for this four-particle
operator contains averages of six-particle operators, and so on. This is a
manifestation of the well-known hierarchy problem of many-particle physics.
A consistent truncation scheme is the cluster expansion~\cite{Fricke1996},
where all occurring operator expectation values are represented by possible
factorizations plus correlations. In our particular case, we use
\begin{eqnarray}
\langle e_\alpha^{\dagger} e_\nu h_\alpha^{\dagger} h_\nu
\rangle & = & \langle e_\alpha^{\dagger} e_\nu\rangle \langle h_\alpha^{\dagger} h_\nu
\rangle\delta_{\alpha\nu} + \delta \langle e_\alpha^{\dagger} e_\nu h_\alpha^{\dagger} h_\nu
\rangle \nonumber \\
& = & f_\nu^{e} f_\nu^{h}\delta_{\alpha\nu} + C^x_{\alpha\nu\alpha\nu} \ ,
\end{eqnarray}
where $C^x_{\alpha\nu\alpha\nu} = \delta \langle e_\alpha^{\dagger} e_\nu
h_\alpha^{\dagger} h_\nu\rangle$ is a measure of
how strongly the electron-hole pairs are correlated. In the cluster
expansion method equations of motion for the correlation
contributions are derived. Then the hierarchy of correlation contributions is
truncated rather than the hierarchy of expectation values itself. This allows
for the consistent inclusion of correlations in the equations of motion
up to a certain order in all of the appearing operator expectation values.

For the following analysis, the equations of motion for the carrier
populations are further evaluated by assuming a temporally
slowly varying interband photon-assisted polarization, so that its adiabatic
solution can be used. This leads to \cite{Jahnke2006,Bayer2006}
\begin{eqnarray}
\frac{d}{dt} f_\nu^{(e,h)}\Big|_{\mbox{\footnotesize opt}} = - \frac{f_\nu^{e} f_\nu^{h}+\sum_\alpha C^x_{\alpha\nu\alpha\nu}}{\tau_\nu},
\label{decayform}
\end{eqnarray}
with the Wigner-Wei{\ss}kopf decay rate
\begin{eqnarray}
\frac{1}{\tau_\nu} = - \frac{2}{\hbar} \lim_{\Gamma\to 0^+}\mbox{Re} \sum_{\xi}
\frac{i|g_{\xi\nu}|^2}{\hbar\omega_\nu^{e}+\hbar\omega_\nu^{h}-\hbar\omega_{\xi}-i\Gamma}
\ .
\end{eqnarray}

\begin{figure}
  \centering
\centerline{\psfig{figure=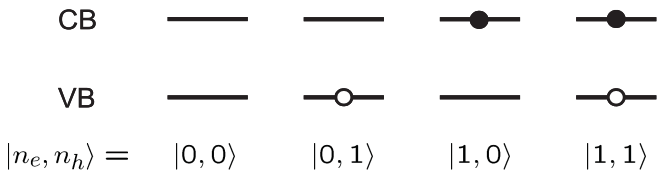,width=\columnwidth}}
\caption{Possible carrier configurations in the conduction and
valence band QD ground states. The spin degree of freedom is
neglected.} \label{fig1}
\end{figure}

In order to illustrate the underlying physics, we consider in the
next two paragraphs only s-shell populations and one spin degree
of freedom of the carriers. The carrier configuration can then be
expanded into the basis set $|n_e, n_h \rangle$, where the $n_e$
and $n_h$ give the number of electrons and holes, respectively
(the photonic part of the states is of no relevance here and not
shown). The possible configurations are $|0,0 \rangle$, $|0,1
\rangle$, $|1,0 \rangle$ and $|1,1 \rangle$, as displayed
schematically in Figure~\ref{fig1}.

If the electron and hole populations were fully correlated, only
$|0,0\rangle$ and $|1,1\rangle$ out of these 4 configurations
would be relevant. Using the following relations for the electron
and hole number operators $e^\dagger e|0,0\rangle = h^\dagger
h|0,0\rangle = 0$ and $e^\dagger e|1,1\rangle = h^\dagger
h|1,1\rangle = |1,1\rangle$ we see that in this {\it two-level
case} $\langle e^{\dagger} e h^{\dagger} h \rangle$ reduces to
$f^e = \langle e^{\dagger} e \rangle$ and also $f^h = \langle
h^{\dagger} h \rangle$. In this particular situation, the source
term of spontaneous emission $f^ef^h+C^x = \langle e^{\dagger} e
h^{\dagger} h \rangle$ in Eq.~(\ref{decayform}) can be replaced by
$f^{(e,h)}$, and then the equations of motion give a
single-exponential decay. As soon as the other two configurations
are included, Eq.~(\ref{decayform}) in general results in a
non-exponential decay. In Section~\ref{sec:numerics} we evaluate
$C^x$ under more general conditions.

\section{Samples and Experiment}
\label{sec:samples} The experiments were performed on different
types of self-assembled (In,Ga)As/GaAs QD arrays fabricated by
molecular beam epitaxy. All samples contained 20 layers of QDs,
which were separated from one another by 60-nm-wide barriers. The
first type of QDs was nominally undoped, the other two types were
modulation doped, one of n-type and the other one of p-type. The
Silicon- or Carbon-doping layers were located 20 nm below each dot
layer. The dopant density was chosen about equal to the dot
density in each layer, so that an average occupation by a single
electron or hole per dot can be expected.

The photoluminescence emissions of the as-grown QD samples are
located around 1200 nm at cryogenic temperatures for all three dot
types. In order to vary the confinement potential, several pieces
from each QD sample type were thermally annealed for 30 s at
different temperatures $T_{ann}$ between 800 and 980 $^{\circ}$C.
Because of the annealing the confinement is reduced due to
intermixing of dot and barrier material. Typical photoluminescence
spectra of the nominally undoped samples, which show the
established behavior for such a series of annealed QD structures
can be found in Ref.~\cite{Greilich2006}. Increasing $T_{ann}$
results in a blueshift as well as a narrowing of the emission line
from the ground state exciton. The corresponding blue shift of the
wetting layer is found to be rather weak as compared to that of
the QD emission. Therefore the confinement potential, which we
define as the energy separation between the wetting layer emission
and the QD ground state emission, varies systematically within an
annealing series. The confinement energies increase from about 50
up to 400 meV with decreasing $T_{ann}$.

The QD samples were mounted on the cold finger of a microscopy
flow-cryostat which allows for temperature variations down to 6K.
In the TRPL studies a mode-locked Ti-sapphire laser emitting
linearly polarized pulses with a duration of about 1 ps at 75.6
MHz repetition rate (corresponding to 13.2 ns pulse separation)
was used for optical excitation. The QD luminescence was dispersed
by a monochromator with 0.5 m focal length and detected by a
streak camera with a S1 photocathode. In the standard synchroscan
configuration, time ranges up to 2 ns could be scanned with a
resolution of about 20 ps. Longer time ranges could be addressed
by installing a long delay time unit of about 50 ps. The excitation was kept as weak as possible to
avoid multiexciton effects.

In the TRDT studies two synchronized Ti-sapphire lasers with a
jitter well below 1 ps were used for the excitation. The emission
energies could be varied independently. One laser beam, the pump,
was used for the creation of carrier populations while the other
one, the probe, was used to test them. The temporal delay between
both pulses could be varied by a mechanical delay line, along
which the probe beam was sent. The transmission of the probe was
detected with a homodyne technique based on phase-sensitive
balanced detection. The polarization of the pump and the probe
beam were chosen either linear or circular co-polarized.

We mention already here that the main topic of our studies is not
the quantitative values of the decay times, which have been
addressed already in many other studies. The focus is instead to
develop a systematic picture of the dependence of the
recombination on experimental parameters, both the internal QD
properties and the external conditions such as excitation energy
and intensity.

\section{Results and Discussion}

\label{sec:discussion} The outline of the carrier recombination
dynamics in Section~\ref{sec:theory} provides a guide for the
experimental studies. An exponential decay could occur if the
carrier populations were correlated, i.e., excitonic
correlations were present. However, in experiments, in which the
carriers are created by non-resonant excitation into the wetting
layer or the barrier, electrons and holes typically relax
independently towards their QD ground states. In this evolution of
the carrier population, dephasing due to carrier scattering
competes with the necessary built-up of excitonic correlations. It
has been discussed for quantum wells in~\cite{hoyer2003} that the
formation process might take longer than the recombination
process. For QDs it has been shown in~\cite{Jahnke2006} that,
while electrons and holes are still localized by the strong
confinement potential, excitonic correlations are easily
suppressed by dephasing processes related to carrier scattering.

In general, the analysis leading to Eq.~(\ref{decayform}) has shown
that the recombination dynamics is determined by (i) the
electron and hole populations, and (ii) the Coulomb
correlations between the carriers. The high flexibility in
fabricating self-assembled QDs allows us to tailor the corresponding
parameters such that their impact can be systematically tested. In
detail, the following experiments have been performed:

(i) The electron and hole populations have been varied by studying
the carrier dynamics in undoped QDs in comparison to those in
either n-type or p-type doped QDs.

(ii) Coulomb interaction can lead to carrier scattering between
QD shells. The carrier scattering can be enhanced by reducing the
shell splitting. Therefore the influence of correlations has been
studied by addressing dots with different confinement heights.

(iii) The correlations can affect carrier scattering only if
enough excess energy is available to fulfil energy conservation
in the scattering event. This excess energy can be varied by the
photon energy of the exciting laser.

\subsection{Influence of excitation energy}

\begin{figure}
  \centering
\centerline{\psfig{figure=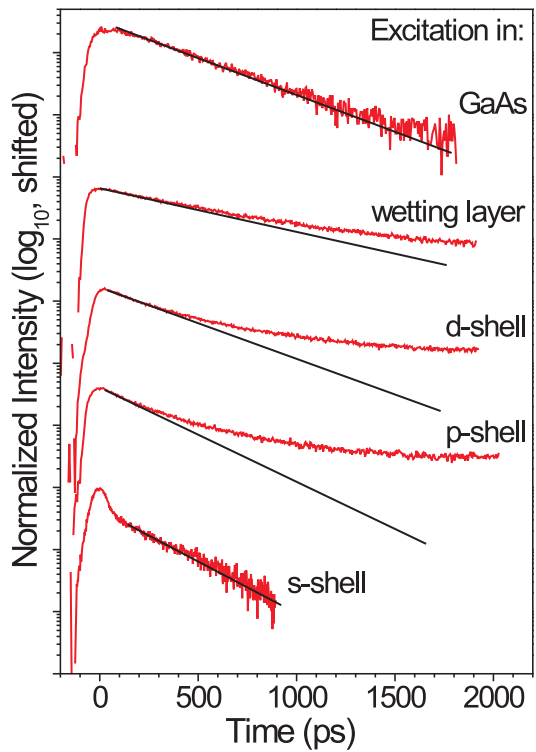,width=7cm}} \caption{TRPL
transients of undoped (In,Ga)As/GaAs self-assembled QDs with a
confinement potential height of 80 meV. Pulsed excitation occurred
at time zero. Detected was the ground state luminescence. The
energy position of the exciting laser is indicated at each trace.
The lines are attempts to fit the data in a time range of 300 ps
after the transient starts to show a clear decay. For resonant
excitation the signal is influenced by scattered laser light
around time zero. $T = 10$~K. The energy of the exciting laser for GaAs, wetting layer, d-shell, p-shell, s-shell has been set to  1.550~eV, 1.476~eV, 1.436~eV, 1.414~eV, 1.389~eV respectively. The average excitation density was 0.7 kW/cm$^2$.}  \label{fig2}
\end{figure}

First we discuss the influence of the available excess energy on
the exciton recombination dynamics. For that purpose, the
excitation was decreased from being non-resonant into the GaAs
barrier to being into the wetting layer, and further into the
confined QD states. Figure~\ref{fig2} shows transients of the
electron-hole recombination from the ground state of nominally
undoped (In,Ga)As/GaAs QDs with a confinement potential of about
80 meV, i.e. the confinement potential in these dots is rather
shallow. The excitation pulse hit the sample at time zero. Note
the logarithmic scale on the left scale.

The top trace shows the result for the GaAs excitation. After a
typical rise of the signal during a few tens of ps, the intensity
drops on a few hundred ps time scale. The solid line shows an
attempt to fit a mono-exponential decay to the data at early
times. For the fit the first 300 ps after the PL plateau maximum
have been used, in this case from 200 to 500 ps. After about a
nanosecond, a clear deviation from this decay can be seen, as
expected from our theoretical model. This deviation becomes more
pronounced for wetting layer excitation, for which already after
700 ps the non-exponential behavior of the decay becomes obvious.
Note further that the rise time of the signal is reduced as
compared to the case of GaAs excitation.

The non-exponential decay is also seen if the excitation is done
below the barrier into the d-shell or the p-shell of the QDs, as
demonstrated by the two mid traces. It has become even more
pronounced than for above barrier illumination, as the deviation
becomes apparent already at earlier delays below 500 ps. At these
delays the decay appears to be faster which might be related to a
more rapid relaxation into the ground state.

Note that these results for below barrier excitation also show
that the deviation from exponentiality cannot be traced to dark
excitons, whose radiative decay requires a spin-flip first. As
soon as carriers are trapped in the QDs, spin relaxation is
strongly suppressed at low $T$, in particular because the
spin-orbit coupling mechanisms which are very efficient in higher
dimensional systems are strongly suppressed.
\cite{BrandesPRB02,KhaetskiiPRL02}. The resulting flip times are
in the microseconds range and may even reach milliseconds, which
is by far too long to give any significant contribution to the
decay dynamics in the monitored time range.

This is consistent with previous observation that the exciton
spin-flip time exceeds tens of ns \cite{PaillardPRL01}. In the
experiment here with a 75.6 MHz laser repetition rate a dark
exciton contribution would appear as constant background at the
low temperatures applied. This is confirmed in studies where the
laser repetition rate was reduced: A slowly decaying background
appears in these experiments for delays exceeding 10 ns, at which
all recombination processes involving optically active excitons
took place.

\begin{figure}
  \centering
\centerline{\psfig{figure=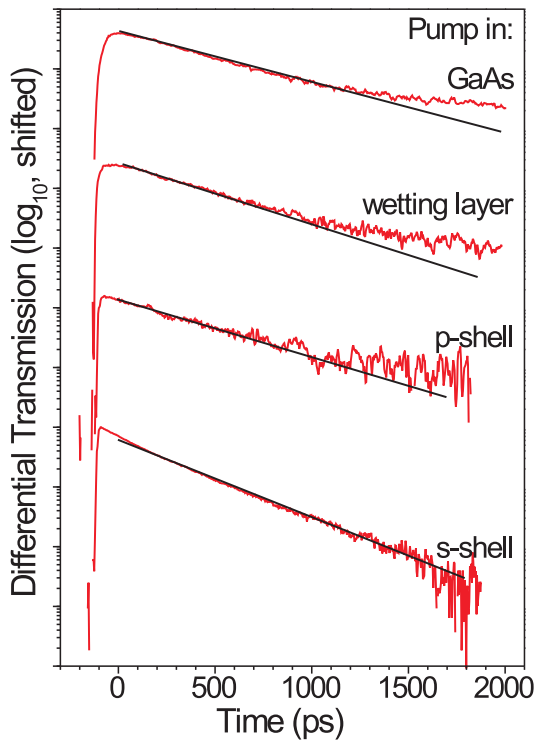,width=7cm}} \caption{TRDT
transients of the (In,Ga)As/GaAs self-assembled QDs with a
confinement potential height of 80 meV studied also in
Fig.~\ref{fig2}. The ground state populations were probed for
different excitation energies of the pump laser as indicated at
each trace. The lines are linear fits to the data in the time
range from 200 to 500 ps. $T$ = 10 K. The energies of the pump laser for GaAs, wetting layer, d-shell, p-shell, s-shell were the same as in the TRPL experiments. The average pump (probe) density was 0.07 kW/cm$^2$ (0.007 kW/cm$^2$).} \label{fig3}
\end{figure}

Varying the excitation power in the regime where multiexciton
effects are negligible leads also to slight variations of the
decay dynamics: For non-resonant excitation
the decay tends to be slowed down in the range of 10\%, while for excitation
into higher lying QD states the changes are weak. For
non-resonant excitation, the deceleration might be attributed to
enhanced carrier diffusion before carrier trapping can occur. For
carrier-carrier scattering which additionally supports the
phonon-assisted relaxation. These observations generally
complicate the interpretation of decay times determined under
non-resonant conditions as exciton lifetimes and, in particular,
the comparison for different samples, as long as the change does
not lie outside of the observed variation range.

The bottom trace of Fig.~\ref{fig2}, finally, shows the TRPL for
resonant excitation between the valence and conduction band ground
states. Around zero delay scattered light from the laser is seen.
After $\sim$ 50 ps a decay becomes prominent, which is within the
experimental accuracy purely exponential, in contrast to the
previous non-resonant excitation conditions. Furthermore, the
decay is much faster than before. Comparing the decay time to
those determined by fitting the early delay data under
non-resonant conditions, we find an acceleration by a factor of
about 2. For non-resonant excitation the optically excited
polarization is converted into populations by the scattering
involved in the relaxation. For resonant excitation, on the other
hand, the carrier coherence is maintained until recombination
occurs, as recent four-wave-mixing studies have demonstrated
\cite{Borri2004}. Therefore under these conditions coherent
luminescence is observed. Corresponding calculations are very
involved as they require additional inclusion of interband
coherence terms in the dynamics. However, from the theory in the
incoherent regime we expect strong carrier correlations in the
case of resonant excitation, i.e. for the source term of
spontaneous emission we have $f^ef^h+C^x \approx f^e$. Hence,
since $f^e > f^ef^h$, Eq.~(\ref{decayform}) predicts a faster
decay for resonant excitation.

The TRPL results are confirmed by TRDT studies shown in
Fig.~\ref{fig3}. The energy of the pump beam was tuned in the same
way as in the TRPL studies described above. The energy of the
probe was fixed to the s-shell. The shape of the different traces
is very similar to those observed in TRPL. For excitation into
GaAs the transmission clearly deviates from an exponential decay,
and the same is true for excitation into the wetting layer, the
d-shell (not shown, very similar to the p-shell case) and the
p-shell. In contrast, for resonant excitation an exponential decay
is observed again with a characteristic time significantly shorter
than that for non-resonant excitation.\\
Under these conditions the exponential decay constants are 310 ps for the TRDT experiment and 280 ps in the case of the TRPL measurement.

\begin{figure}
  \centering
\centerline{\psfig{figure=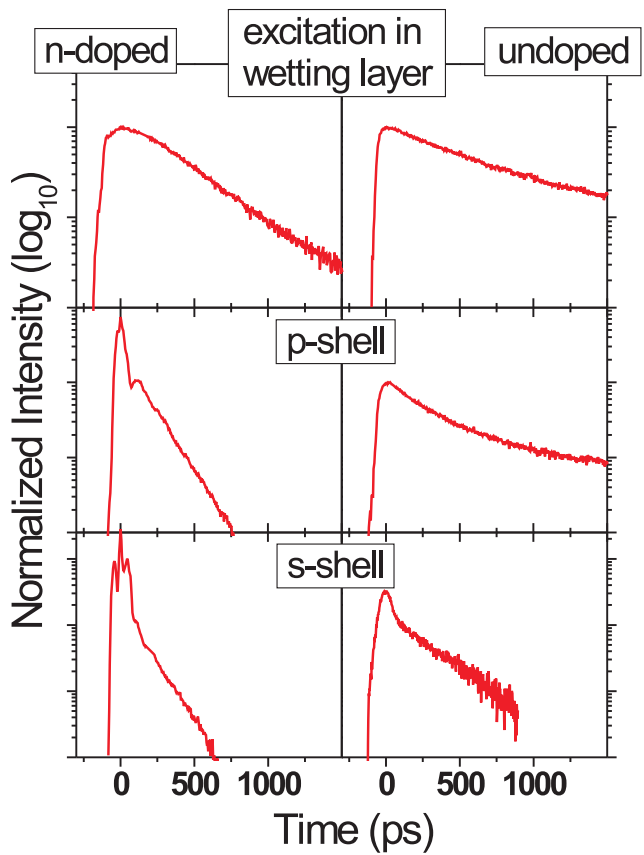,width=8.5cm}}
\caption{Comparison of TRPL traces for n-doped (left panels) and
undoped (right panels) QDs excited at different energies, as
labelled in each figure. The height of the confinement potential
is about $80\,$meV. Around time zero the signal is influenced by
scattered laser light. $T$ = 10 K. The exitation energy for the n-doped sample in GaAs, wetting layer, d-shell, p-shell, s-shell has been set to 1.550 eV, 1.476 eV, 1.437 eV, 1.417 eV, 1.397 eV respectively. The average excitation density was 0.7 kW/cm$^2$.}
\label{fig4}
\end{figure}

\subsection{Influence of doping}

Neglecting the influence of Coulomb correlations, according to
Eq.~(\ref{decayform}) the carrier population dynamics can be
pushed towards a mono-exponential decay if either the electron or
the hole population is approximately held constant. This can be
achieved by a background doping, for which we studied both n- and
p-doped samples which were prepared such that there is on average
a single carrier per dot. The studies show that besides variations
in the quantitative values for the decay times the shape is very
similar, independent of the type of doping. Therefore we focus on
the n-doped structures only.

Figure~\ref{fig4} depicts the corresponding TRPL results for
n-doped QDs, excited at different energies. The confinement
potential was about 80 meV. For comparison the data for the
undoped dots from Fig.~\ref{fig2} are also shown. Clearly, the
decay behavior of the doped dots comes much closer to an
exponential decay, independent of the actual excitation energy.
Again, only for resonant excitation, however, mono-exponential
decays are seen in both cases.

For non-resonant excitation such as in GaAs also the n-doped QDs
show a deviation from an exponential decay at long delays. While
this might be well correlated with the influence of correlation
induced scattering, we cannot exclude some contribution from
charge neutral QDs, where the charge depletion might partly arise
from above barrier photoexcitation.

We note that these results give also some hint why the PL decay in
the undoped QDs is closer to an exponential behavior for
excitation into GaAs than for wetting layer excitation. It is well
known that non-resonant excitation into the barrier may lead to a
formation of charged excitons, for which the decay in Fig.
\ref{fig4} is almost exponential. Formation of charged complexes
is strongly suppressed for  below barrier excitation in undoped
QDs.

\subsection{Influence of correlations}

\begin{figure}
  \centering
\centerline{\psfig{figure=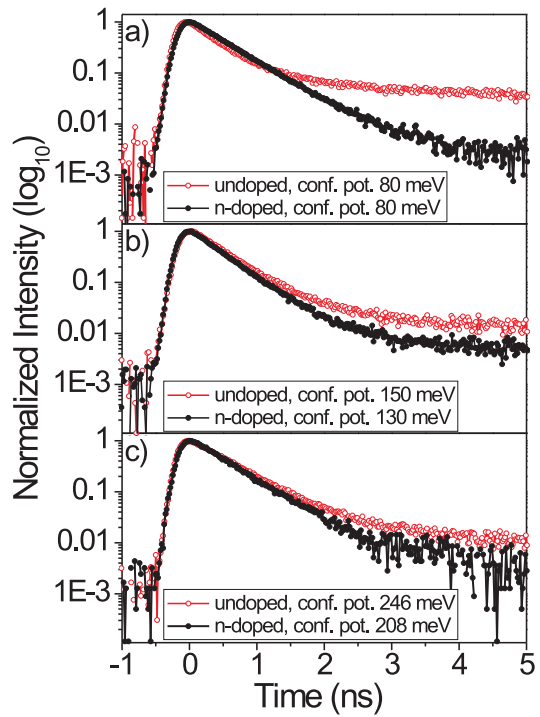,width=7cm}} \caption{TRPL
transients for undoped (open symbols) and n-doped (full symbols)
QDs with different confinement potentials, as indicated in each
panel. Excitation was done into GaAs at 1.550 eV. $T$ = 10 K, excitation density 0.7 kW/cm$^2$.} \label{fig5}
\end{figure}

The magnitude of the correlations between carriers due to Coulomb
interaction can be tailored by varying the QD confinement. With
increasing confinement potential the splitting between the dot
shells increases, while possible scattering (that suppresses
correlations) is reduced. This was studied by comparing QDs
annealed at different temperatures. Figure~\ref{fig5} shows the
results for QD samples which were excited non-resonantly into
GaAs. For comparison, again the data for undoped and n-doped QDs
are displayed. The height of the confinement potentials increased
from 80 (80) to 150 (130) and further to 250 (210) meV for undoped
(n-doped) structures. The resulting splittings between the
confined QD shells, as estimated from high excitation PL
spectroscopy, are 20, 35, and 50 meV, respectively.

In all cases it can be seen that the dynamics in the undoped dots
deviates more strongly from an exponential decay than that in the
doped structures. However, with increasing confinement the
difference becomes smaller, and for the strongest confinement the
traces almost coincide. In this particular case the influence of
the Coulomb scattering has been reduced to an extent that it is no
longer relevant for the dephasing of correlations.

\section{Numerical results}

\label{sec:numerics} In this section we provide exemplary
numerical results which support the previous conclusions. The
semiconductor luminescence equations (SLE) are used to describe
the time evolution of the photon number $\langle b_\xi^\dagger
b_\xi\rangle$, the carrier populations $f_\nu^{(e,h)}$, the
photon-assisted polarization $\langle b_{\xi}^{\dagger} h_\nu
e_\nu \rangle$, and the carrier-carrier correlations such as
$C^x_{\alpha\nu\alpha\nu} = \delta \langle e_\alpha^{\dagger}
e_\nu h_\alpha^{\dagger} h_\nu\rangle$. Scattering is treated in
relaxation-time approximation. We restrict ourselves to the
formulation of the theory in the incoherent regime, as presented
in~\cite{Jahnke2006}, and consider nonresonant excitation. The QD
parameters are those used in Ref.~\cite{Jahnke2006}, except the QD
density is $N = 10^{10}\;\mbox{cm}^{-2}$, the dipole moment is
$16.8e\mathring{A}$ and the dephasing of the correlations is
$0.05\,$meV. Even though the dephasing is weak it effectively
destroys the correlations on a time scale of tens of ps.

\begin{figure}
  \centering
\centerline{\psfig{figure=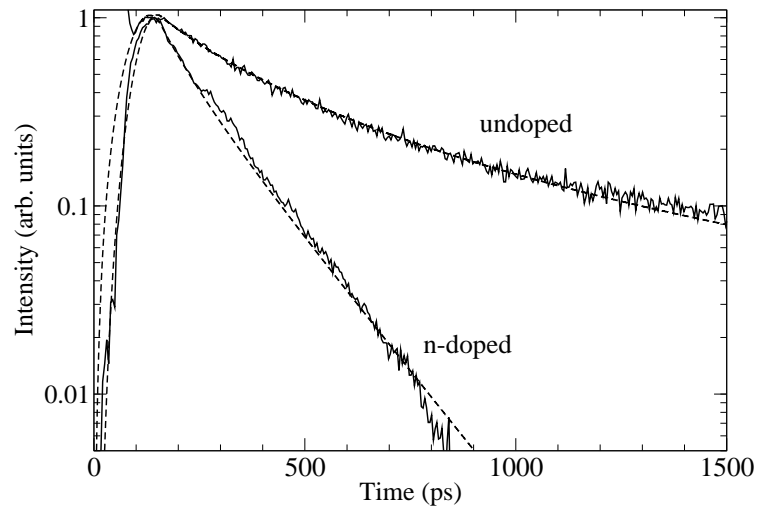,width=8cm}}
\caption{Calculated TRPL intensity (dashed lines) according to
Eq.~(\ref{intensity}) for pumping into the p-shell of undoped and
n-doped QDs. The experimental data (solid lines) are the same as
in Fig.~\ref{fig4}.} \label{fig6}
\end{figure}

Figure~\ref{fig6} shows results for undoped and n-doped QDs
excited in the p-shell. For the undoped situation we pump the
system with equal electron and hole density $N_e=N_h=0.35N$. In
the n-doped case we assume on average one additional electron per
QD, i.e. $N_e = N_h+N$ with again $N_h = 0.35N$. Apart from this
difference in the initial conditions both curves have been
calculated with exactly the same parameters. An agreement between
theory and experiment can be observed: (i) the doped QDs show an
exponential decay, whereas the undoped ones show a non-exponential
decay. (ii) the decay is much faster for the doped QDs if compared
to the undoped QDs.

\begin{figure}
  \centering
\centerline{\psfig{figure=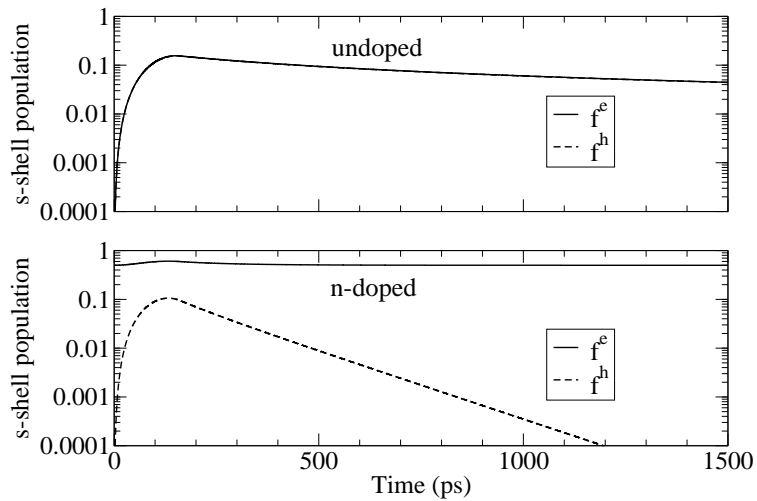,width=8cm}} \caption{Time evolution of
  electron and hole populations, $f^e$ and $f^h$ in the s-shell of undoped (top) and n-doped (bottom)
QDs. The population is defined such that it is unity if the
s-shell is populated by two carriers with opposite spin. Single
carrier population corresponds to a 0.5 population accordingly.}
\label{fig7}
\end{figure}

To understand the origin of these different behaviors, it is
illuminating to study the time evolution of the s-shell
populations as depicted in Fig.~\ref{fig7} for one spin subsystem.
In the undoped case the s-shell populations are zero at first. Due
to the pump process and the subsequent carrier scattering, the
s-shell population increases temporarily and decays subsequently
to its initial value. In the n-doped case the electron occupation
in the s-shell starts with the finite value of 0.5 due to the
doping. The temporal change of the electron population relative to
the doping level is small. According to Eq.~(\ref{decayform}), a
constant electron population $f^e_\alpha$ leads to an exponential
decay of the hole population $f^h_\alpha$ and, hence, of the
PL-intensity for the considered situation of strong suppression of
excitonic correlations $C^x$ due to dephasing.

\section{Conclusions}

In summary, we have performed a detailed study of the carrier
recombination dynamics in QDs. The results show that the carrier
recombination in general follows a non-exponential decay. Only
under specific conditions, like resonant excitation, strong
confinement, or intentional doping, a mono-exponential decay is
observed. In addition, ensuring coherence of the excited carriers
by resonant excitation leads to a strong shortening of the decay
time. The experimental results are in excellent agreement with
numerical results obtained from a microscopic theory which
abandons the shortcomings of the commonly used two-level
description of QDs.

{\bf Acknowledgements.} We gratefully acknowledge the financial
support of this work by the Deutsche Forschungsgemeinschaft
(research group `Quantum Optics in Semiconductor Nanostructures'
and the reseach project BA 1549/10-1). The Bremen group
acknowledges a grant for CPU time at the NIC, Forschungszentrum
J\"ulich.


\end{document}